\documentclass[12pt]{article}
\usepackage[utf8]{inputenc}
\usepackage{graphicx} 
\usepackage{dcolumn} 
\usepackage{bm} 
\usepackage{amsfonts}
\usepackage{amsthm}
\usepackage{amsmath}
\usepackage{amssymb}
\usepackage{epsfig}
\usepackage{color}
\usepackage{textcomp}
\usepackage{hyperref}
\usepackage{titlesec}
\usepackage{slashed}
\usepackage{caption}
\usepackage{subcaption}
\usepackage{booktabs}
\usepackage{multirow}
\usepackage{cite}
\usepackage{placeins}
\usepackage{comment}
\usepackage{braket}
\usepackage{authblk}
\usepackage{physics}

\let\a=\alpha \let\b=\beta  \let\g=\gamma  \let\d=\delta \let\e=\varepsilon
  \let\h=\eta     
\let\m=\mu    \let\n=\nu             
     
   \let\o=\omega

  \let\del=\nabla

\def\\{\hfill\break} \let\==\equiv

\let\0=\noindent

\let\dpr=\partial

\def\qed{\hfill\raise1pt\hbox{\vrule height5pt width5pt depth0pt}}

\def\be{\begin{equation}}
\def\ee{\end{equation}}
\def\bea{\begin{eqnarray}}\def\eea{\end{eqnarray}}
\renewcommand{\theequation}{\arabic{section}.\arabic{equation}}
\begin{document}

\title{A Simple Derivation of the Gertsenshtein Effect}

\author[1]{Andrea Palessandro \thanks{apalessandro@deloitte.dk}}
\author[2]{Tony Rothman \thanks{tonyrothman@gmail.com}}
\affil[1]{\small Deloitte Consulting, Artificial Intelligence and Data}
\affil[2]{\small New York University, Department of Applied Physics (retired)}

\date{}

\maketitle

\begin{abstract}
\noindent As shown by Gertsenshtein in 1961, an external magnetic field can catalyze the mixing of graviton and photon states in a manner analogous to neutrino-flavor oscillations.  We first present a straightforward derivation of the mechanism by a method based on unpublished notes of Freeman Dyson.  We next extend his method to include boundary conditions and retrieve the results of Boccaletti et al. from 1970. We point out that, although the coupling between the graviton and photon states is extremely weak, the large magnetic fields around neutron stars $\sim 10^{14}$ G make the Gertsenshtein effect a plausible source of gravitons. We also point out that axion-photon mixing, a  subject of active current research, is essentially the same process as the Gertsenshtein effect, and so the general mechanism may be of broad astrophysical and cosmological interest.

\end{abstract}
\section{Introduction}
\setcounter{equation}{0}\label{sec1}
\baselineskip 8mm

In a classic 1961 paper \cite{Gert61}, Mikhail Gertsenshtein demonstrated that electromagnetic waves passing through an external magnetic field in curved space could be transmuted into gravitational waves. The result was entirely classical but in quantum mechanical language one can say that the external field ``catalyzes" a resonant mixing between photon and graviton states in a manner analogous to the mixing of neutrino flavors. Although the coupling between the graviton and photon states is extremely weak, the large magnetic fields associated with, for example, neutron stars, make the Gertsenshtein effect a plausible source for producing gravitons \cite{RB06}.  Zel'dovich \cite{Zel73}, however, and Dyson \cite{Dyson05a,Dyson05b} pointed out that the Gertsenshtein mechanism requires coherence between the gravitational and electromagnetic waves. Large magnetic fields  create electron-positron pairs from the vacuum, which in turn alters its index of refraction. The speed of electromagnetic wave propagation is lowered relative to that of the gravitational wave and the Gertsenshtein mechanism is quenched.

The Gertsenstein effect received some attention in the 1970s when Boccaletti et al. \cite{Bocc70} provided a comprehensive discussion and de Logi and Mikelson \cite{deLogi77} calculated the cross section for graviton-photon conversion.  Cross sections have also been computed by Long et al. \cite{Long94}, and the effect has been discussed by Ejlli \cite{Ejlli13} as well as Jones and Singleton \cite{Jones18}. More recently, the mechanism has been mentioned in the context of the primordial gravitational wave, or graviton, background by Vagnozzi and Loeb \cite{Vagnozzi:2022qmc} and various others \cite{add1,add4,add5,add6,add7}. Nevertheless, most physicists apparently remain unaware of the Gertsenshtein effect, although  current calculations of axion-photon conversion employ a nearly identical formalism (see \S 4). 
 
As it turns out, Freeman Dyson rediscovered the Gertsenshtein effect in 2005 \cite{Dyson05a}, evidently  unaware of the previous work. When informed that he was not first, he chose not to publish his notes, but gave them to one of us (TR). As Dyson's derivation is simpler than the others we have seen and has remained unpublished, we thought it worthwhile to make it available in some form. In the following section of this paper we present Dyson's approach, filling in some steps and  clarifying a few ambiguities.  In \S 3 we  make contact with Boccaletti et al.'s results.  It will be seen that Dyson's method is much simpler. Finally in \S 4 we point out the similarity to axion-photon conversion, and conclude by indicating some possible applications of the Gertsenshtein effect in the cosmological context.

\section{Graviton-Photon Oscillations}
\setcounter{equation}{0}\label{sec2}
Consider the propagation of electromagnetic and gravitational waves in a flat spacetime.  For the gravitational wave (GW),  we assume as usual that it is represented by a weak perturbation $h_{\m\n}({\bf x}, t) \ll 1$ traveling on a Minkowski background such that the full spacetime metric is
\be
g_{\m\n} = \eta_{\m\n} + h_{\m\n} \ \ ;\ \  g^{\m\n} = \eta^{\m\n} - h^{\m\n}\ \ ;\ \   h^{\m\n} \equiv
\h^{\a\m}\h^{\b\m}h_{\a\b}, \label{gmunu}
\ee
where we take the flat-space metric to be $\h_{\m\n} = (-1,1,1,1)$.

We carry out our calculations in the standard transverse-traceless (TT), or Lorentz, gauge. If the wave is taken to be propagating in the $z$-direction, then $h_{\m\n} = h_{\m\n}(z,t)$ and the transverse components of the field are $h_{11} = -h_{22}$ and $h_{12} = h_{21}$, which represent the two independent polarizations of the GW.\footnote{We use units in which $G = c = 1$ and follow MTW \cite{MTW73} conventions throughout. In particular, Greek indices $= 0...3$ are spacetime indices, while Latin indices are spatial indices $= 1...3$. Repeated indices in any position are summed.}

In the Lorentz gauge, the linearized gravitational wave equation (see \cite{MTW73}, chap.18 and \cite{Weinberg72}, chap.10) is
\be
\Box h_{\mu\nu} = -16\pi T_{\mu \nu},\label{boxh}
\ee
with $\Box \equiv \dpr^\m \dpr_\m$. We assume that the energy-momentum tensor $T_{\m\n}$ consists entirely of the electromagnetic energy of the system. In doing so, we ignore gravitational backreaction, which is a good approximation as long as $h_{\m \n} \ll 1$. Therefore we have
\be
4\pi T_{\m\n} = F_{\m\a}F_\n\,^\a -\frac1{4}g_{\m\n}F_{\a\b}F^{\a\b}, \label{Tuv}
\ee
where $F_{\m\n} \equiv  \frac{\dpr A_\n}{\dpr x^\m} - \frac{\dpr A_\m}{\dpr x^\n}$ is the usual electromagnetic field tensor and $A_\m = (-\phi, \textbf{A})$ is the  vector potential. In terms of the electric and magnetic fields, the Maxwell stress tensor reads
\be
4\pi T_{ij}= -(E_iE_j + B_iB_j)+\frac1{2}\d_{ij}(\bf E \cdot E + B \cdot B).\label{Tij}
\ee
With our conventions, $F_{ij} = A_{j,i} - A_{i,j} = \e_{ijk}B_k$ and $F_{i0} = A_{0,i} - A_{i,0} \equiv - \partial_i \phi - \dot{A}_i = E_i$.\footnote{Because $h_{11} = -h_{22}$, we must also have $T_{11} = -T_{22}$, which is not generally true  for the stress tensor (\ref{Tij}) unless $E_3 = B_3 = 0$. This will be the case for transverse electromagnetic waves traveling in the $z$-direction (``radiation gauge"), which we also assume.} (Here, $\e_{ijk}$ is the usual permutation operator.)

We further assume that the electromagnetic wave obeys the vacuum Einstein-Maxwell equations:
\be
\frac1{\sqrt g}(\sqrt g F^{\a\b}),_\b = 0 . \label{EM}
\ee
Since with metric (\ref{gmunu}) the determinant is $g = 1 + {\cal O}(h_{ij}^2) \approx 1$,(\ref{EM}) amounts to
\be
(g^{\a\m}g^{\b\n}F_{\m\n}),_\b = 0.
\ee
Working to first order in $h_{\m\n}$ we have
\be
\h^{\a\m} \h^{\b\n} F_{\m\n},_\b - \h^{\a\m}h^{\b\n}F_{\m\n},_\b -\h^{\a\m}F_{\m\n}h^{\b\n},_\b -\h^{\b\n}h^{\a\m}F_{\m\n},_\b -\h^{\b\n}F_{\m\n}h^{\a\m},_\b = 0.
\ee
However, in the Lorentz gauge $h^{\b\n},_\b \equiv 0$, so the third term vanishes. Discarding the second and fourth terms as small, but keeping the final term as potentially large leaves
\be
\h^{\a\m}\h^{\b\n}F_{\m\n},_\b - \h^{\b\n}h^{\a\m},_\b F_{\m\n} = 0.\label{EM2}
\ee
For $\a = 0$  (\ref{EM2}) gives merely ${\bf \del \cdot E} = 0$, as in flat space. For $\a = i$, the equation   becomes
\be
-\dot E_i + \e_{ik\ell}B_{\ell},_k + \dot h_{ij}E_j -h_{ij},_k\e_{jk\ell}B_\ell = 0. \label{EM3}
\ee
(Since $h^{ij} = h_{ij}$, it is immaterial whether one writes this equation with indices up or down.) The $\dot E_i$ term can be eliminated as in freshman physics  to get a wave equation for $B$.  Taking the curl of  (\ref{EM3}) yields
\be
-\e_{rpi}\dot E_i,_p + \e_{rpi}\e_{ik\ell}B_\ell ,_{kp}+\e_{rpi}( {\dot h_{ij}E_j),_p - \e_{rpi}\e_{jk\ell}(h_{ij},_k B_\ell}),_p = 0. \label{EM4}
\ee
By virtue of the homogeneous Maxwell equation $\bf \del \times E = -\dot B,$ which is the same as in flat space, the first term is $\dpr^2 B_r/\dpr t^2 \equiv B_{r,tt}$. By virtue of the homogeneous Maxwell equation $\bf \del \cdot B = 0$, the second term is $\curl\curl \textbf{B} = -\nabla^2 \textbf{B} \equiv - B_{r,\ell \ell}$.\footnote{Here we have used the  vector  identity $\curl \curl \textbf{A} = \nabla (\nabla \cdot \textbf{A}) -\nabla^2 \textbf{A}$}

Next, we assume that the electric field $\bf E$ is entirely that of the electromagnetic wave, but that the magnetic field consists of the field $\bf b$ of the electromagnetic wave plus a  background field ${\bf B}_0$, such that ${\bf B = B}_0 + {\bf b}$
with ${\bf B}_0 \gg \bf b$; ${\bf B}_0 \gg \bf E $. Thus we can ignore the third term above as second-order small and   (\ref{EM4}) becomes
\be
 B_r,_{tt} -B_r,_{\ell\ell} - \e_{rpi}\e_{jk\ell}(h_{ij},_{k}B_\ell),_p = 0. \label{EM6}
\ee

With the same assumptions as above and discarding terms $\sim b^2$, the gravitational wave equation (\ref{boxh}) becomes
\be
\Box h_{ij} = 4[B_{0i}B_{0j} + B_{0i}b_j + b_iB_{0j} -\frac{1}{2} \d_{ij}(B_{0i}^2 + 2 B_{0k}b_k)].\label{boxh2}
\ee

The trick is, first, to choose ${\bf B}_0 = \text{constant}$,\footnote{Note  that for $i=j=3$,  (\ref{boxh2}) gives $\Box h_{33} \sim \textbf{B}_0^2 \neq 0$. This entails the appearance of a longitudinal component of the gravitational wave $h_{33}$, induced by $T_{33}$. Given that in our case $\Box h_{33} \sim {\bf B}_0^2 = \text{constant}$, this mode does not mix with the other two transverse components; therefore, for the purpose of our derivation, we can effectively ignore it.} which linearizes both (\ref{EM6}) and (\ref{boxh2}), and secondly, to choose  ${\bf B}_0$ at right angles to $\bf b$, which ensures a nonzero coupling for the $h_{12}$ mode in  (\ref{boxh2}). Assuming that the electromagnetic wave is traveling in the $z$-direction, we take ${\bf b = b}_y(z,t)$  and   ${\bf B}_0 \equiv {\bf B}_{0x}$.  Then from   (\ref{EM6}):
\be
 \ddot b_y - b_y'' = h_{12}''B_{0},\label{EM7}
\ee
where the prime ($'$) indicates differentiation with respect to $z$, and from  (\ref{boxh2}):
\be
\ddot h_{12} - h_{12}'' = - 4B_0b_y.\label{boxh3}
\ee
Note that the background field ${\bf B}_0$ makes the mixing between a vector field (photon) and a tensor field (graviton) possible by providing the missing angular momentum. Given that $\Box h_{ij} \sim T_{ij} \sim E_i E_j + B_i B_j$, for the given field configuration the only possible linear mixing is between the $h_{12}$ component of the gravitational wave and the $b$-field of the electromagnetic wave. There is no interaction with the other polarization state $h_{11}$.\footnote{Another way to see this is to note that the polarization tensors for the GW can be written as the dyadics ${\bf e}_+ = \bf \hat i\hat i-\hat j\hat j$ and ${\bf e}_\times = \bf\hat i\hat j+ \hat j\hat i$.  Thus, ${\bf e}_+ \cdot {\bf B}_0 \ || \ {\bf B}_0 $ and ${\bf e}_+ \cdot {\bf b} \ || \ {\bf b}$, so the $h_{11}$ mode produces no interaction between ${\bf B}_0$ and ${\bf b}$, whereas ${\bf e}_\times \cdot {\bf B}_0 \ || \ {\bf b}$ and ${\bf e}_\times \cdot {\bf b} \ || \ {\bf B}_0$, so the $h_{12}$ mode does allow interaction between ${\bf B}_0$ and ${\bf b}$.}

It  remains now only to solve the two coupled equations (\ref{EM7}) and (\ref{boxh3}). We search for ``wave packet" solutions $f$ the form $ f(z,t) = f_r(z-t)f_s(z+t)$, where $r$ stands for ``rapid" and $s$ stands for ``slow". The Ur-example would be
\bea
f &=& \cos(k_r z - \o_r t)\cos(k_s z + \o_s t)\nonumber \\&=& \frac{1}{2}\left[\cos((k_r + k_s)z + (\o_s - \o_r) t)+ \cos((k_r - k_s)z - (\o_r + \o_s) t)\right],\nonumber
\eea
where $\o_r \gg \o_s$ and in our units $k_r = \o_r \, ; k_s = \o_s$. Such wave packets can thus be regarded either as product functions consisting of a high and low frequency component or as a coherent superposition of waves traveling in the same direction, which can be expected to form beats as in elementary physics.

Dropping subscripts, we write $h= h_r(z-t)h_s(z+t)$ and $b= b_r(z-t)b_s(z+t)$, where in our units $h_r' = -\dot h_r$ and $h_s' = \dot h_s$, with similar expressions for $b_r$ and $b_s$.  Inserting the trial functions into (\ref{EM7}) and (\ref{boxh3}) gives
\be
b'_rb'_s = -\frac{B_0}{4} (h''_rh_s + 2 h'_rh'_s + h_rh''_s) \approx -\frac{B_0}{4}h''_rh_s,\label{EM8}
\ee
and
\be
h'_rh'_s = B_0b_rb_s\label{boxh4}.
\ee
In  (\ref{EM8}) we have approximated $h''$ as $h_r''h_s$, since by assumption $h_r$ is rapidly varying and $h_s$ is slowly varying.

The equations can now be solved by inspection. Pick
\be\label{hbeq}
h'_r  = \frac{\sqrt{B_0}}{a} b_r\ \ ; \ \ h'_s = a\sqrt{B_0}b_s  \ \ ; \ \ b'_s=-\frac{B_0^{3/2}}{4a}h_s,
\ee
with $a$ an arbitrary constant. These choices satisfy  (\ref{EM8}) and (\ref{boxh4}). The functions $b_r$ and $h_r$ which describe the shape of the wave-packets can also be chosen arbitrarily and have no effect on the behavior of the functions $b_s$ and $h_s$, which describe the slow oscillation of the packets between photon and graviton states. 

Letting $a = 1/2$ gives
\be\label{hbeq2}
h'_s = \frac{1}{2} \sqrt{B_0} b_s  \ \ ; \ \ b'_s = -\frac{1}{2} B_0^{3/2} h_s,
\ee
which have solutions
\be
h_s = {\cal A}\sin\left(\frac{z+t+\phi}{L}\right)\ \  ;\ \ b_s = {\cal A} \sqrt{B_0} \cos\left(\frac{z+t+\phi}{L}\right),
\label{GEsol}
\ee
provided that $L \equiv 2/B_0$. Notice also that 
\be
h_s^2 + \frac{b_s^2}{B_0} ={\cal A}^2=\text{constant}, \label{A2}
\ee
which shows that the total energy remains constant as the packet oscillates (see the appendix for details). The frequency of oscillation is given by the magnitude of the magnetic field: $\omega_s \equiv L^{-1} = B_0/2$.

We have seen that a uniform classical magnetic field $\textbf{B}_0$ catalyzes a linear mixing of the gravitational and photon fields. A quantum of the mixed field oscillates between photon and graviton states with a mixing length $L$ independent of wavelength, namely
\be
L = \frac{2}{B_0} \label{mixingL},
\ee
where $B_0$ is the component of the background field perpendicular to the direction of propagation of the quantum. This means that if a single photon travels a distance $D$ through the uniform field $B_0$, it will emerge as a graviton with probability
\be
P = \sin^2(D/L) \label{prob}.
\ee
The quadratic dependence of $P$ on $D$ makes this process interesting as a possible astrophysical source of gravitons.

In cgs units 
\be
L = \frac{2 c^2}{\sqrt{G} B_0} \approx 2 \text{Mpc} \left(\frac{1 \,\text{Gauss}}{B_0}\right).
\ee
Around a magnetar, the field can be $\sim 10^{14}$ G, leading to a mixing length of only $\sim 10^6$ km.  From (\ref{prob}) above, an observer stationed within the boundaries of such a field at a distance $D \gg L$ from the star could in principle observe a significant periodic change of magnetar luminosity on a timescale of seconds. More realistically, the extent of a neutron star's magnetic field is $\sim 10$ km. Beyond its field boundaries oscillations cease (\S 3) and the brightness of the star is diminished by a factor of $10^{-10}$ in the direction perpendicular to the magnetic field. Conceivably, two observers could infer the presence of the Gertsenstein mechanism by comparing photon fluxes in directions parallel and perpendicular to the neutron star $B$-field. However, since the effect is classical, such an observation would not constitute a direct detection of gravitons \cite{RB06, BR06, Palessandro:2019tmj}.

We also note that $10^{14}$ G is approximately the Schwinger limit, at which one expects the electromagnetic field equations to become nonlinear and electron-positron pairs to be created.  Pair creation would in turn alter the index of refraction of the vacuum and change the speed of light, destroying the presumed coherence of the gravitational and electromagnetic waves. Hence, for such strong fields one should do a more detailed calculation before drawing definite conclusions.

\section{Inclusion of Boundary Conditions}
\setcounter{equation}{0}\label{sec3}

In the previous section we obtained a general solution for graviton-photon oscillations. Boccaletti et al. \cite{Bocc70} employed a more conventional equation-solving method to obtain a solution in which conversion takes place within a finite region, say for $0 \le z \le D$.  Such a scenario is probably more relevant to graviton production in the vicinity of neutron stars, for example.

We now employ the Dyson approach to retrieve their results for photon-to-graviton conversion in this situation.

We assume that the constant background magnetic field $\textbf{B}_0 \equiv \textbf{B}_{0x}$ is confined to the finite region  $0 \le z \le D$ (region II), with a plane right-moving electromagnetic wave traveling along the $z$-axis incident upon it. By the Gertsenshtein effect, the electromagnetic wave produces a left-traveling gravitational wave in region I ($z<0$), and a right-traveling gravitational wave in region III ($z>D$).

In the region where the magnetic field is non-zero, the transverse component of the gravitational wave $h_{12}$ is described by a wave packet of the form\footnote{As we saw in \S 2, the function $h_r(z-t)$ can be chosen arbitrarily. Here we make the simplifying assumption that $h_r(z-t) \propto \sin(\omega_r(z-t))$.}
\begin{equation}\label{packet}
h_s(z+t) h_r(z-t) = \mathcal{A} \sin \left( \omega_s (z+t) \right) \sin \left(\omega_r (z-t)\right)
\end{equation}
where $\omega_r$ ($\omega_s \equiv L^{-1}$) is the frequency of the rapidly (slowly) oscillating component of the metric, $\cal A$ is the maximum amplitude of the wave, and for simplicity we have set the global phase $\phi=0$ in (\ref{GEsol}).

The most general solution for the gravitational wave in region II  is then a superposition of wave packets of the form (\ref{packet}) and plane waves. Following \cite{Bocc70}, we turn now to a complex formalism to describe these waves, with the  understanding that physical solutions are always given by the real part of complex amplitudes.

Using this formalism, $h_{12}$ in region II is given by (the real part of)
\begin{equation}\label{h12II}
    h_{12}^{II}(z,t) = a e^{i\omega_r(z-t)} + b e^{-i\omega_r(z+t)} + \mathcal{A}e^{-i\omega_s(z+t)} e^{i\omega_r(z-t)} \quad \text{for} \quad 0 \le z \le D.
\end{equation}
The first two terms represent right-traveling and left-traveling plane waves respectively, while the third term is a superposition of wave packets.\footnote{Specifically, the third term is a combination of the four elementary wave packets $\cos(\omega_s(z+t))\cos(\omega_r(z-t))$, $\cos(\omega_s(z+t))\sin(\omega_r(z-t))$, $\sin(\omega_s(z+t))\cos(\omega_r(z-t))$, and $\sin(\omega_s(z+t))\sin(\omega_r(z-t))$.}

In both regions $z<0$ and $z>D$ where the background magnetic field is zero, the wave equation for the metric is $\Box h_{1 2} = 0$, and the corresponding vacuum solutions are
\be
h_{12}^{\text{I}}(z,t) = c e^{-i\omega_r(z+t)} \quad \text{for} \quad z<0,
\ee
and
\be
h_{12}^{\text{III}}(z,t) = d e^{i\omega_r(z-t)} \quad \text{for} \quad z>D,
\ee
describing a generic left-traveling ($z<0$) or right-traveling ($z>D$) plane wave. 

In order to determine the constant coefficients $(a,b,c,d)$ in terms of $\mathcal{A}$ we impose the usual boundary conditions demanding continuity of the functions and their derivatives at $z = 0$ and $z=D$: 
\begin{equation}\label{boundary}
    \begin{split}
        h_{12}^{\text{I}}(0,t) = h_{12}^{\text{II}}(0,t), \\
        \partial_z h_{12}^{\text{I}}(0,t) = \partial_z h_{12}^{\text{II}}(0,t), \\
        h_{12}^{\text{II}}(D,t) = h_{12}^{\text{III}}(D,t), \\
        \partial_z h_{12}^{\text{II}}(D,t) = \partial_z h_{12}^{\text{III}}(D,t).
    \end{split}
\end{equation}
Or, in terms of the coefficients:
\begin{equation}\label{boundary2}
    \begin{split}
    c = a + b + \mathcal{A} e^{-i\omega_st}, \\
    c = b - a - \left(1-\frac{\omega_s}{\omega_r}\right)\mathcal{A} e^{-i\omega_s t},\\
    d = a + b e^{-2i\omega_rD} + \mathcal{A} e^{-i\omega_s(D+t)}, \\
    d = a - b e^{-2i\omega_rD} + \mathcal{A} e^{-i\omega_s(D+t)} \left(1-\frac{\omega_s}{\omega_r}\right).
    \end{split}
\end{equation}
This is a system of four linear equations for five unknowns. In terms of $\mathcal{A}$, coefficients $(a,b,c,d)$ are:
\begin{equation}\label{consts}
    \begin{split}
        a = - \mathcal{A} \left(1-\frac{\omega_s}{2 \omega_r}\right) e^{-i \omega_s t},\\
        b = - \mathcal{A} \frac{\omega_s}{2 \omega_r} e^{iD(2\omega_r - \omega_s)} e^{- i \omega_s t}, \\
        c = \mathcal{A} \frac{\omega_s}{2\omega_r}\left(1-e^{iD(2\omega_r-\omega_s)}\right)e^{-i\omega_s t}, \\
        d = - \mathcal{A} \left(1 - \frac{\omega_s}{2\omega_r}\right)(1-e^{-iD\omega_s})e^{-i\omega_st}.
    \end{split}
\end{equation}
The presence of the time-dependent factor $\exp(-i\omega_st)$ means that the coefficients in (\ref{consts}) are not truly constant. However, given our assumption that $e^{-i\omega_s t}$ is slowly varying, to first approximation we can consider them time-independent:
\begin{equation}\label{constsapprox}
    \begin{split}
        a = - \mathcal{A} \left(1-\frac{\omega_s}{2 \omega_r}\right),\\
        b = - \mathcal{A} \frac{\omega_s}{2 \omega_r} e^{iD(2\omega_r-\omega_s)}, \\
        c = \mathcal{A} \frac{\omega_s}{2\omega_r}(1-e^{iD(2\omega_r -\omega_s)}), \\
        d = - \mathcal{A} \left(1-\frac{\omega_s}{2\omega_r}\right)(1-e^{-iD\omega_s}).
    \end{split}
\end{equation}
One can easily check that the metric $h_{12}$ with the coefficients in (\ref{constsapprox}) satisfies the wave equation in all three regions (in our approximation) as well as the boundary conditions in (\ref{boundary}). Hence, the full solution for $h_{12}(z,t)$ is
\begin{equation}\label{hfull}
    \begin{split}
        h_{12}^I(z,t) = \mathcal{A} \frac{\omega_s}{2\omega_r}(1-e^{iD(2\omega_r-\omega_s)}) e^{-i\omega_r(z+t)}, \\
        h_{12}^{II}(z,t) = - \mathcal{A} \left(1-\frac{\omega_s}{2\omega_r}\right) e^{i\omega_r(z-t)} - \mathcal{A} \frac{\omega_s}{2\omega_r} e^{iD(2\omega_r-\omega_s)}e^{-i\omega_r(z+t)} \\
        + \mathcal{A} e^{-i\omega_s(z+t)}e^{i\omega_r(z-t)}, \\
        h_{12}^{III}(z,t) = - \mathcal{A} \left(1-\frac{\omega_s}{2\omega_r}\right)(1-e^{-iD\omega_s}) e^{i\omega_r(z-t)}.
    \end{split}
\end{equation}

We can similarly derive a solution for the electromagnetic wave. From equation (\ref{hbeq}), $b_r = h_r'/(2\sqrt{B_0})$ and $b_s = 2 h_s' / \sqrt{B_0}$. Given that $h_r(z-t) = e^{i \omega_r(z-t)}$ and $h_s(z+t) = \mathcal{A} e^{-i\omega_s(z+t)}$, we have $b_r(z-t) = i \omega_r/(2\sqrt{B_0}) e^{i\omega_r(z-t)}$ and $b_s(z+t) = - 2 i \mathcal{A} \omega_s/\sqrt{B_0} e^{-i\omega_s(z+t)}$, so the electromagnetic wave packet in region II is of the form
\begin{equation}\label{bII}
    b_r(z-t) b_s(z+t) = \frac{1}{2} \mathcal{A} \omega_r e^{-i\omega_s(z+t)} e^{i\omega_r (z-t)}.
\end{equation}
By assumption, the incident electromagnetic wave is right-moving in both regions I and III. In addition, we need to include a produced left-moving electromagnetic wave in region I, as well as a combination of right-moving and left-moving plane waves in region II, as we did for the gravitational wave in (\ref{h12II}). Consequently, after imposing the relevant boundary conditions as in (\ref{boundary}) and disregarding the slowly oscillating term $\exp(-i\omega_st) \approx 1$, the full solution for $b(z,t)$ is:
\begin{equation}\label{bfull}
    \begin{split}
        b^I(z,t) = \mathcal{B} \omega_r e^{i \omega_r(z-t)} + \frac{1}{4} (1-e^{iD(2\omega_r-\omega_s)}) \mathcal{A} \omega_s e^{-i\omega_r(z+t)},\\
        b^{II}(z,t) = \left[\mathcal{B} - \frac{1}{2} \mathcal{A} \left( 1-\frac{\omega_s}{2\omega_r} \right)\right] \omega_r e^{i\omega_r(z-t)} - \frac{1}{4} e^{iD(2\omega_r - \omega_s)} \mathcal{A} \omega_s e^{-i\omega_r(z+t)} \\
        + \frac{1}{2}\mathcal{A} \omega_r e^{-i\omega_s(z+t)} e^{i\omega_r (z-t)}, \\
        b^{III}(z,t) = \left[ \mathcal{B} - \frac{1}{2} \mathcal{A}\left(1 - \frac{\omega_s}{2\omega_r} \right) (1-e^{-iD\omega_s}) \right] \omega_r e^{i\omega_r(z-t)},
    \end{split}
\end{equation}
where $\mathcal{B}$ is the incident electromagnetic wave amplitude. Thus far, the coefficient $\mathcal{B}$ is undetermined because we have had to solve for five coefficients but have only four boundary conditions.  However, one can easily check that the wave equation (\ref{boxh3}) for $h_{12}$ in region II is  consistent with the second of (\ref {bfull}) only if $\mathcal{B} = \mathcal{A}/2$.  One can also check that the resulting expressions for $h_{12}$ and $b$ conserve energy.

The solutions (\ref{hfull}) and (\ref{bfull}) describe the conversion of an electromagnetic wave to a gravitational wave, catalyzed by the constant background magnetic field $B_0$. In region II where $B_0 \neq 0$, the electromagnetic and gravitational amplitudes slowly oscillate between each other via the conversion term $\exp(-i\omega_s(z+t))$. Outside, in regions I and III, the oscillation stops and the wave packets propagate independently.

The solution of \cite{Bocc70} is an approximation of ours for $\omega_s \ll \omega_r$. In fact, to leading order in $\omega_s$, the gravitational wave solution in (\ref{hfull}) becomes
\begin{equation}\label{happrox}
    \begin{split}
        h_{12}^I(z,t) = \mathcal{A} \frac{\omega_s}{2\omega_r}(1-e^{2iD\omega_r}) e^{-i\omega_r(z+t)}, \\
        h_{12}^{II}(z,t) = \mathcal{A} \frac{\omega_s}{2\omega_r} e^{i\omega_r(z-t)} - \mathcal{A} \frac{\omega_s}{2\omega_r} e^{2iD\omega_r}e^{-i\omega_r(z+t)} - i \omega_s z \mathcal{A} e^{i\omega_r(z-t)}, \\
        h_{12}^{III}(z,t) = - i D \omega_s \mathcal{A} e^{i\omega_r(z-t)},
    \end{split}
\end{equation}
where we have used  $e^{-i\omega_s z} \approx 1 - i \omega_s z$  for small $\omega_s$.

The coefficients of (\ref{happrox}) are the same as in \cite{Bocc70}, given that the authors use a convention in which the wave packet solution is given by $\lambda z/(2i\omega_r) \exp(i\omega_r(z-t))$, with $\lambda = 2 \omega_r \omega_s \mathcal{A}$. In this approximation, the amplitude of the gravitational wave in region I vanishes if the condition $\exp(2iD\omega_r) = 1$ is satisfied, and the amplitude of the gravitational wave in region III depends linearly upon the length and magnitude of the static field.

Similarly, the approximate solution for the electromagnetic wave is, to leading order in $\omega_s$,
\begin{equation}\label{bapprox}
    \begin{split}
        b^I(z,t) = \frac{1}{2} \mathcal{A} \omega_r e^{i \omega_r(z-t)} + \frac{1}{4} (1-e^{2iD\omega_r}) \mathcal{A} \omega_s e^{-i\omega_r(z+t)},\\
        b^{II}(z,t) = \frac{1}{2} \mathcal{A} \omega_r e^{i\omega_r(z-t)}  - \frac{1}{4} e^{2iD\omega_r} \mathcal{A} \omega_s e^{-i\omega_r(z+t)} - \frac{1}{2} i \mathcal{A} \omega_r \omega_s z e^{i\omega_r (z-t)}, \\
        b^{III}(z,t) = \frac{1}{2} \left(1 - i D \omega_s \right) \mathcal{A} \omega_r e^{i\omega_r(z-t)}.
    \end{split}
\end{equation}
If the condition $e^{2iD\omega_r} = 1$ is satisfied, the back-propagating wave amplitude vanishes and only the incident wave is left in region I. The solution in this approximation describes an incident electromagnetic wave of amplitude $\mathcal{A}$ and energy $\omega_r$ that propagates into region II and slowly converts to a gravitational wave with a rate given by $\omega_s \equiv L^{-1}$. When the wave enters region III, the incident electromagnetic wave amplitude is effectively reduced by a factor of $i D \omega_s$ while the gravitational wave amplitude is increased by that same factor:
\begin{align}
 \begin{split} 
        &\text{EM}: \quad \mathcal{A} \rightarrow \mathcal{A} \left(1 - i \frac{D}{L}\right), \\ \
        &\text{GW}: \quad  0 \rightarrow - i \mathcal{A} \frac{D}{L}.
\end{split}
\end{align}

\section{Axion-Photon Conversion}
\setcounter{equation}{0}\label{sec4}
Quite generally the Lagrangian density for the interaction of gravitational waves with matter is
\be
{\cal L} = -\frac12 h_{\m\n}T^{\m\n} ,
\ee
where in our problem $T^{\m\n}$ is given by (\ref{Tuv}) and (\ref{Tij}).

Currently, there is a large interest in axion-photon conversion \cite{Anast17, Adair22, Matthews22}, as axions are presently the leading candidates for dark matter. In a process much like the Gertsenshtein mechanism, axions emerging from the sun can be transformed into X-ray photons by a strong laboratory magnetic field and subsequently detected by an X-ray telescope. Indeed, the axion-photon interaction Lagrangian can be written as
\be
{\cal L} = -\frac{g_{a\g}}{4} a F_{\m\n}\Tilde{F}^{\m\n},
\ee
where $\Tilde{F}^{\m\n} \equiv \frac12 \e^{\m\n\a\b}F_{\a\b}$ is the dual electromagnetic field tensor, $a$ is the axion amplitude and $g_{a\g}$ is the axion-photon coupling constant. 

In terms of the electric and magnetic fields, the above expression amounts to ${\cal L}= g_{a\g}\, a \, {\bf E \cdot B}$, which is essentially the same Lagrangian as the one we have been using except that the external magnetic field couples to the $E$-field of the photon rather than the $B$-field.

Not surprisingly, then, we can  employ Dyson's approach for axion-photon mixing. In the high-energy limit in which one can ignore the mass of the axion, and in the presence of a background magnetic field $B_0$, the coupled equations for the axion $a$ and the E-field of the photon $e$ are analogous to (\ref{EM7}) and (\ref{boxh3}):
\begin{equation}
    \begin{split}
        \ddot{a} - a'' = - 4 g_{a \g} B_0 e \\
        \ddot{e} - e'' = g_{a \g} a'' B_0,
    \end{split}
\end{equation}
and they admit oscillatory solutions with mixing length
\begin{equation}
    L = \frac{2}{g_{a \g} B_0},
\end{equation}
which agrees with the high-energy limit of \cite{Anast17}.  As one can see, this is for all intents and purposes the result of \S 2.

\section{Conclusions}
The Gertsenshtein effect describes the conversion of electromagnetic waves into gravitational waves and vice-versa.  Conversion is made possible by an external magnetic field that provides the extra angular momentum necessary for a spin-1 field to mix with a spin-2 field.

In this paper, we have presented a novel approach to deriving the Gertsenshtein effect due to Freeman Dyson. The key step in his method is to decompose the simultaneous solution of the linearized gravitational wave and Einstein-Maxwell equations into rapidly and slowly oscillating components. The rapidly oscillating component describes the ``shape" of the electromagnetic and gravitational wave packets, while the slowly oscillating component describes the ``beats" between the two fields, with the resulting oscillations having a mixing length independent of wavelength. 

In \S 3 we extended Dyson's method to include boundary conditions and retrieved the result of \cite{Bocc70} as a special case of our solution. We then employed (\S 4) the same method in the context of axion-photon conversion with essentially identical results. Given the quadratic form of Lagrangians in general and the similarity of the axion-photon Lagrangian to the gravitational interaction Lagrangian in particular, we may conjecture that Gertshenstein-like mechanisms which cause mixing among multiple fields are ``universal" and have widespread cosmological utility. Our approach can almost certainly be applied to any scenario involving these mechanisms. 

In future papers we plan to extend the effect to the (original) Yang-Mills SU(2) field as well as consider the possibility that graviton-photon mixing in the long-wavelength limit might generate something resembling a cosmological constant.\\

\noindent {\bf Acknowledgement:}
Knowing that any expression of gratitude will fall far short, TR would like to thank the late Freeman Dyson for forty years of friendship and inspiration.  Among his many acts of generosity, one of the minor ones was to make available the notes on which this project was based. 

\newpage
\appendix
\section{Conservation of Energy}
\renewcommand{\theequation}{A.\arabic{equation}}
\setcounter{equation}{0}
In this appendix we demonstrate in detail the statement made in \S 2, that energy is conserved during the oscillations between the electromagnetic and gravitational fields.

Choosing for simplicity $h_r(z-t) = \sin(\omega_r(z-t))$ as the rapidly oscillating component of the metric, and retaining factors of $G$, the gravitational wave packet is given by
\begin{equation}\label{hshr}
    h(z,t) = h_s(z+t) h_r(z-t) = \mathcal{A} \sin(\omega_s(z+t)) \sin(\omega_r(z-t)),
\end{equation}
while, by virtue of (\ref{hbeq}), the electromagnetic wave packet is given by
\begin{equation}\label{bsbr}
    b(z,t) = b_s(z+t) b_r(z-t) = \frac{1}{2 \sqrt{G}}\mathcal{A} \omega_r \cos(\omega_s(z+t)) \cos(\omega_r(z-t)).
\end{equation}
Thus the slowly oscillating components are out of phase by $\pi/2$, but the rapidly oscillating components are as well, and so the above equations  do not in fact produce beats. The reason is that (\ref{hshr}) is not the physical gravitational wave packet. 

To see this, note that the energy density of the electromagnetic wave is
\begin{equation}\label{eEM}
    \epsilon_{\text{EM}} = \frac{b^2}{8 \pi} = \frac{1}{32 \pi G} \mathcal{A}^2 \omega_r^2 \cos^2(\omega_s(z+t)) \cos^2(\omega_r(z-t)),
\end{equation}
while the energy density of the gravitational wave is
\begin{equation}\label{eGW}
     \epsilon_{\text{GW}} = \frac{\dot{h}^2}{32 \pi G} \approx \frac{1}{32 \pi G} \mathcal{A}^2 \omega_r^2 \sin^2(\omega_s(z+t)) \cos^2(\omega_r(z-t)),
\end{equation}
where we have approximated $\dot{h} \approx \mathcal{A} \omega_r \sin(\omega_s(z+t)) \cos(\omega_r(z-t))$ given that by assumption $\omega_s \ll \omega_r$. Equations (\ref{eEM}) and (\ref{eGW}) \textit{do} describe coherent oscillations in the energy density of the two fields,  suggesting that we take for the physical solution of the gravitational wave packet
\begin{equation}\label{tildeh}
    \Tilde{h} \equiv \frac{\dot{h}}{2 \sqrt{G}} = \frac{1}{2 \sqrt{G}}\mathcal{A} \omega_r \sin(\omega_s(z+t)) \cos(\omega_r(z-t)).
\end{equation}
Indeed, the wave packets (\ref{bsbr}) and (\ref{tildeh}) describe a right-traveling plane wave of frequency $\omega_r$ that slowly oscillates between gravitational and electromagnetic states with frequency $\omega_s$. Summing (\ref{eEM}) and (\ref{eGW}) and averaging over time gives
\begin{equation}\label{energycons}
    \langle \epsilon_{\text{EM}} + \epsilon_{\text{GW}} \rangle = \frac{1}{8 \pi} \langle b^2 + \Tilde{h}^2 \rangle = \frac{\mathcal{A}^2 \omega_r^2}{64 \pi G},
\end{equation}
showing that the total energy of the system is conserved as the packet oscillates.

\newpage

{\small

\end{document}